# Improved Critical Current Density of $MgB_2$– Carbon Nanotubes (CNTs) Composite

Chandra Shekhar*, Rajiv Giri, S.K. Malik [a] and O N Srivastava*

Department of Physics, Banaras Hindu University, Varanasi-221005, India

[a] Tata Institute of Fundamental Research, Mumbai-400005, India

- *Email address: hepons@yahoo.com ( O. N. Srivastava )
  chand_bhu@yahoo.com ( Chandra Shekhar )

Tel: 0091 542 2368438
Fax: 0091 542 2369889, 2368174

**Abstract**:

In the present study, we report a systematic study of doping/ admixing of carbon nanotubes (CNTs) in different concentrations in $MgB_2$ .The composite material corresponding to $MgB_2$–x at.% CNTs (35 at.% ≥ x ≥ 0 at.%) have been prepared by solid-state reaction at ambient pressure. All the samples in the present investigation have been subjected to structural/ microstructural characterization employing XRD, Scanning electron microscopic (SEM) and Transmission electron microscopic (TEM) techniques. The magnetization measurements were performed by Physical property measurement system (PPMS) and electrical transport measurements have been done by the four-probe technique. The microstructural investigations reveal the formation of $MgB_2$–carbon nanotube composites. A CNT connecting the $MgB_2$ grains may enhance critical current density due to its size (~ 5–20 nm diameter) compatible with coherence length of $MgB_2$ (~ 5–6 nm) and ballistic transport current carrying capability along the tube axis. The transport critical current density ($J_{ct}$) of $MgB_2$ samples with varying CNTs concentration have been found to vary significantly e.g. $J_{ct}$ of the $MgB_2$ sample with 10 at.% CNT addition is ~2.3 x $10^3$ A/cm$^2$ and its value for $MgB_2$ sample without CNT addition is ~7.2x$10^2$ A/cm$^2$ at 20K. In order to study the flux pinning effect of CNTs doping/ admixing in $MgB_2$, the evaluation of intragrain critical current density ($J_c$) has been carried out through magnetic measurements on the fine powdered version of the as synthesized samples. The optimum result on $J_c$ is obtained for 10 at.% CNTs admixed $MgB_2$ sample at 5K, the $J_c$ reaches ~5.2 x$10^6$ A/cm$^2$ in self field, ~1.6 x $10^6$ A/cm$^2$ at 1T, ~2.9 x $10^5$ A/cm$^2$ at 2.6T and ~3.9 x $10^4$ A/cm$^2$ at 4T. The high value of intragrain $J_c$ in 10 at.% CNTs admixed $MgB_2$ superconductor has been attributed to the incorporation of CNTs into the crystal matrix of $MgB_2$, which are capable of providing effective flux pinning centres. A feasible correlation between microstructural features and superconducting properties has been put forward.

**Introduction:**

Doping/ admixing with nano particles has attracted much attention since the discovery of superconductivity at 40K in $MgB_2$ by Akimitsu and co-worker [1]. The unexpectedly high critical temperature of $MgB_2$ is close to the upper limit of the BCS type superconductivity. A conventional phonon mediated pairing mechanism in this material is supported by the observation of a significant boron isotope effect,[2] scanning tunneling experiments [3, 4] and decreased $T_c$ values under hydrostatic pressure [5, 6]. The low mass of boron, leading to high phonon frequency, is thought to be responsible for such a high $T_c$, as supported by band structure and phonon calculations [7, 8]. Recently many researchers have focused their studies on the effect of chemical substitution on the structure and properties of $MgB_2$. As one of the few successful substitutions, C substitution at B site has been carried out by several groups[9–14]. The results on C solubility and the effect of C doping on $T_c$ reported so far vary significantly due to the precursor materials, fabrication techniques and processing condition used. Earlier studies have shown that C does substituted at B site, causing decrease in the 'a' lattice parameter but 'c' lattice parameter remains unchanged. Many materials acted as the carbon precursors, such as amorphous carbon, $B_4C$ and carbon nanotubes (CNTs), all resulted in some improvements in $H_{c2}$ and critical current density. Among various precursors, CNTs exhibited some particularities for their high aspect ratio and nanometer diameter. The effects of carbon nanotubes (CNTs) doping/ addition on $MgB_2$ have been reported [15–19]. In these studies authors have mainly focused their findings on effect of CNTs on $T_c$, and $J_c$. The CNTs doping has been found to improve both vortex pinning and $H_{c2}$, but the underlying pinning mechanism is not known. However, the detailed studies of microstuctural features of $MgB_2$–CNTs composites and their correlation with superconducting properties have not been studied so far.

It is known that microstructural features affect crucially the critical current density of superconducting materials. Therefore, in order to explore the microstructural characteristics and its possible correlation with superconducting properties, particularly $J_c$, in the present paper, we have carried out a systematic study of different doping/ admixing concentration of CNTs in $MgB_2$ prepared by solid-state reaction at ambient pressure.

The transport critical current density ($J_{ct}$) of $MgB_2$ samples with varying CNTs concentration have been found to vary significantly e.g. $J_{ct}$ of $MgB_2$ sample with 10 at.% CNTs addition is ~2.3 x $10^3$ A/cm$^2$ and its value for $MgB_2$ sample without CNTs addition is ~7.2x$10^2$ $A/cm^2$ at 20K. The optimum result on intragrain $J_c$ is obtained for 10 at.% CNTs admixed $MgB_2$ sample at 20K, the $J_c$ reaches ~3.2 x$10^6$ $A/cm^2$ in self field and ~4.0 x $10^5$ $A/cm^2$, ~6.2 x $10^3$ $A/cm^2$, and 1.9 x $10^3$ $A/cm^2$ at 1T, 2.6T and 4T respectively. In the present investigation, we have shown that some of the carbon from CNTs (present as impurities in CNTs sample) has been substituted for boron in $MgB_2$ lattice. On the other hand, CNTs do not decompose, but become a part of the crystal matrix as a whole, where they are effective pinning centers.

## Experimental details

The $MgB_2$–CNTs composites were prepared by conventional solid state reaction. The CNTs used in our experiments were prepared by spray pyrolysis method[20]. Powders of high purity magnesium (99.9%), amorphous boron (99%) and CNTs of ~ 5–20 nm diameter were mixed in acetone medium homogeneously according to nominal atomic ratio of $MgB_2$–x at.% CNT (x = 0, 5, 10, 15, 25 and 35 at.%). The grounded powders were cold pressed (3.0 tons/inch$^2$) into small rectangular pellets (10 x 5 x 1) mm$^3$ and encapsulated in a Mg metal cover to circumvent the formation of MgO during sintering process. The pellet configuration was put on a Ta boat and sintered in flowing Ar atmosphere in a tube furnace at 600$^o$C for 1h, at 800$^o$C for 1h and at 900$^o$C for 2h. Then the pellets were cooled to room temperature at the rate of 100$^o$C/h. The pellet was taken out and encapsulating Mg cover was removed. More details of the synthesis procedure can found from our earlier publication [21]. All the samples in the present investigation have been subjected to gross structural characterization by X-ray diffraction technique (XRD, Philips PW-1710 $CuK_\alpha$), electrical transport characterization by four-probe technique (Keithley resistivity Hall set-up), surface morphological characterization by scanning electron microscopy (SEM, Philips XL20) and the microstructural characterization by transmission electron microscopy (Philips EM-CM-12). The $J_{ct}$ values of all the samples have been measured by standard four probe technique using the criteria of 1$\mu$V/cm. In this measurement, we have made a micro bridge on the bar shape samples and four linear contacts were made with the help of highly conducting silver glue. The magnetization measurements have been carried out at Tata Institute of Fundamental Research (Mumbai, India) over a temperature range of 5-40K employing a physical property measurement system (PPMS, Quantum Design). Intragrain $J_c$ (magnetic $J_c$) was calculated from the height '$\Delta$M' of the magnetization loop (M-H) using Bean's critical state model [22]. In the present investigation magnetization measurements have been carried out on fine ground powders of the samples. In the fine powder form, strong coupling between the grains is non-existent; the intragrain $J_c$ can be estimated employing Bean's formula and using average size of the powder particle. Usually, the particles after grinding of samples may not correspond to singular grains but are as estimated through SEM to be small agglomerates of nearly spherical shape (~10 µm ) covering only few grains. The intragrain critical current density ($J_c$) can be estimated by using Bean's formula:

$$J_c = \frac{30\Delta M}{< d >}$$

where '$\Delta$M' is change in magnetization with increasing and decreasing field (in emu/cm$^3$) and 'd' is average sample size (in cm).

## Results and discussion

Fig.1 shows the representative XRD pattern of the $MgB_2$–x at.%CNT (35 at% ≥ x ≥ 0 at%) superconducting samples. The peaks in the pattern can be well indexed by $MgB_2$ and CNTs. The CNTs can be detected by XRD in the form of graphitic carbon. As higher density of CNTs were mixed in the sample, the intensity of (101) peak of carbon became stronger, indicating the presence of CNTs. The crystal lattice parameters of 10at.% CNTs admixed and pure $MgB_2$ have been found to be a = 0.30581 nm, c = 0.3522 nm and a =0.3083 nm, c = 0.3524 nm respectively. We propose that due to the presence of CNTs and their catalytic influence, the contraction of B–B plane in $MgB_2$ takes place and the lattice parameter 'a' becomes smaller than that of $MgB_2$, while the lattice parameter 'c' remains nearly constant. The decrement in the lattice parameter 'a' implies that carbon atoms (present as impurities in CNTs sample) may have gone into the honeycomb-net plane of boron, resulting in a contraction of the B–B plane.

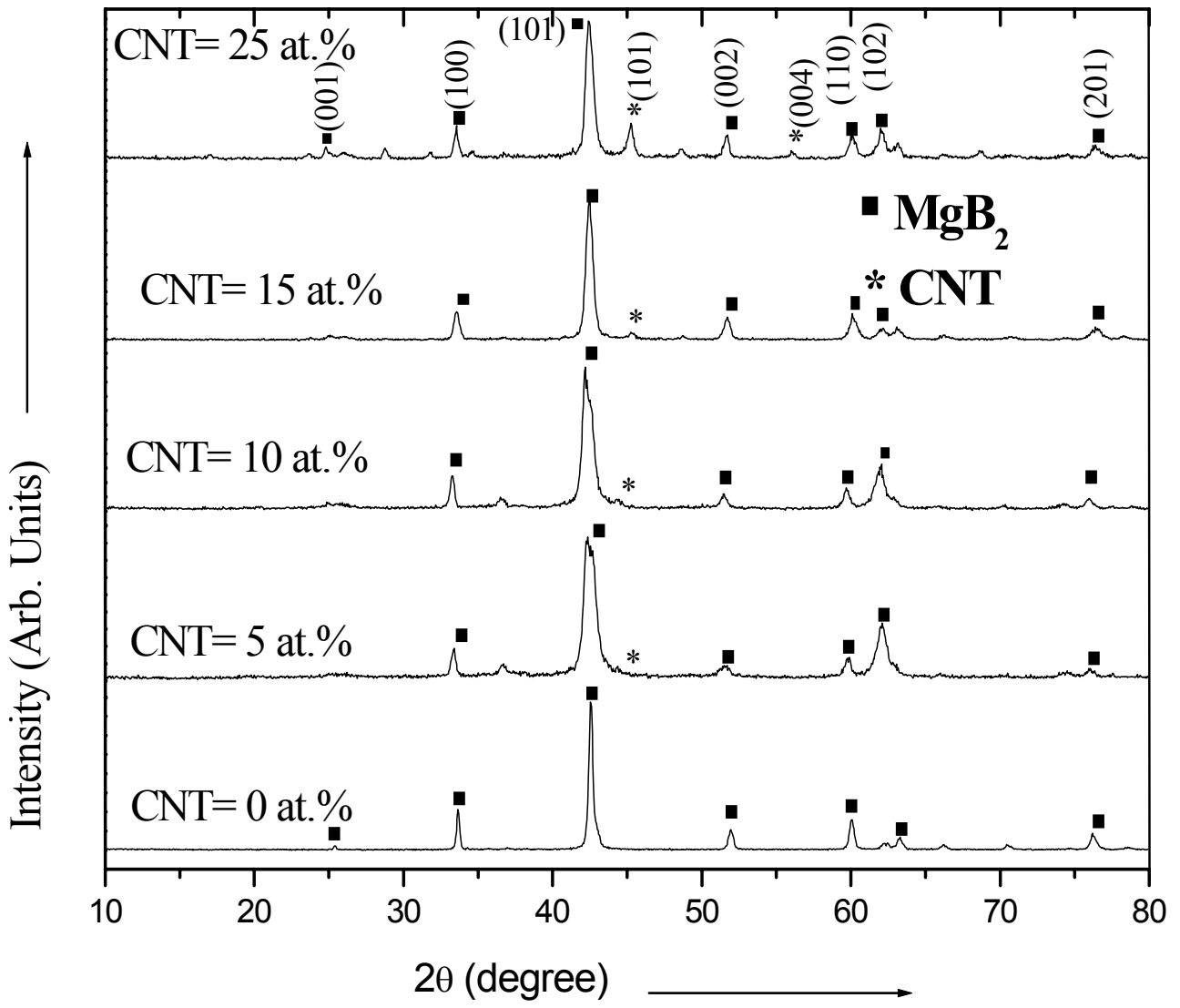

Fig. 1: Representative powder X-ray diffraction patterns of $MgB_2$–x at% CNTs composite samples (x = 0, 5, 10, 15 and 25 at.%)

The variation of resistance with temperature of $MgB_2$–CNTs composites with varying amount of CNTs has been measured by standard four-probe method and is shown in Fig. 2. The transition temperature of the as synthesized $MgB_2$–CNTs composites span a range ~27 to 40K. The transition temperature of $MgB_2$–CNTs samples decreases as the CNTs content increases. Samples show metal like behavior upto the CNTs content $x < 25$ at.% above 40K and semiconducting behaviour when CNTs content exceeds $x \geq 25$ at.%. The inset of Fig.2 shows the dependence of the transition temperature ($T_c$) on the CNTs content. It can be seen that $T_c$ of the samples decreases as CNTs content increases. The decrease in $T_c$, based on the known results can be taken to be due to substitution of C in the honeycomb-net plane of boron. Since carbon is unlikely to come from a very stable configuration like CNT, it seems that the carbon, which is present as impurities in the CNT sample (there was no purifications done for the CNT sample synthesized by us), gets incorporated in the boron network. This is in consistent with the recent reports on carbon doped $MgB_2$ [10, 11, 23]. The increase of resistivity results due to a relative high resistivity of CNTs (the CNTs employed by us have semiconducting characteristics) compared to $MgB_2$.

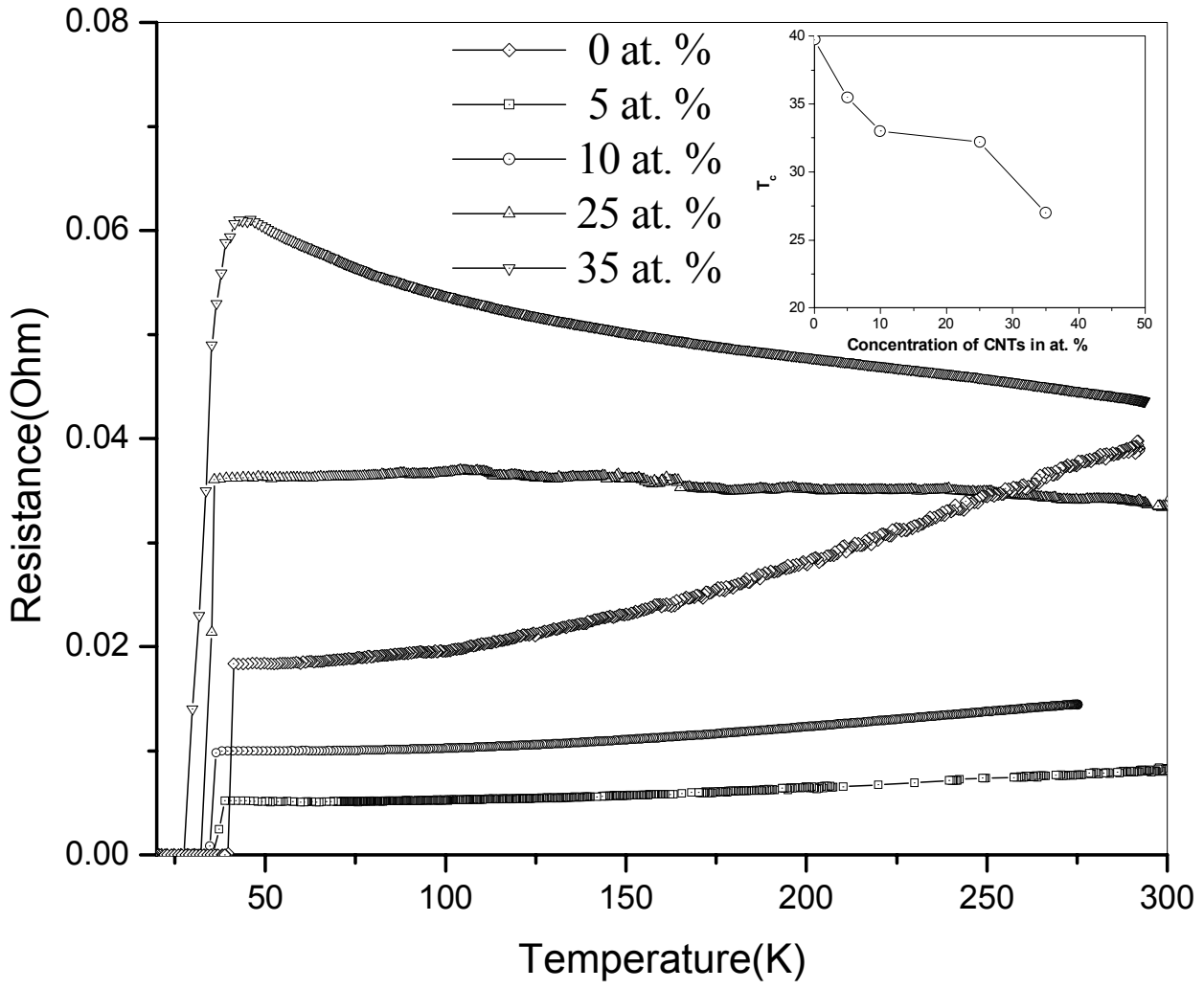

Fig. 2: Resistance vs Temperature behavior of of $MgB_2$–x at% CNTs composite samples (x = 0, 5, 10, 25 and 35 at.%) Inset shows the variation of transition temperature with different of CNTs.

The transport critical current density ($J_{ct}$) of all the samples has been measured by the four-probe technique using criteria of 1μV/cm. The variation of $J_{ct}$ as a function of temperature for 10 at.% CNTs admixed samples sintered at different temperature is shown in Fig.3a. It is clear from Fig.3a that $J_{ct}$ achieves highest value for samples sintered at 900 $^0$C. The $J_{ct}$ as a function of temperature for $MgB_2$–x at.%CNT samples ($0 \leq x \leq 25$ at.%) sintered at 900 $^0$C is shown in Fig.3b The highest $J_{ct}$ value ~2.3 x $10^3$ A/cm$^2$ at 20K has been obtained for 10 at.% CNTs admixed $MgB_2$ sample. The $J_{ct}$ value for 15 at.% and 25 at.% CNTs admixed $MgB_2$ samples are ~1.51 x $10^3$ A/cm$^2$ and ~1.0 x $10^3$ A/cm$^2$ respectively at 20K. However, the $J_{ct}$ value of pure sample is ~0.718 x $10^3$ A/cm$^2$ at 20K.The high value of transport critical current density may occur either due to good connectivity due to admixing of CNTs in $MgB_2$ grains or present of precipitate of secondary particles along the grain boundaries or some unique topological structure connecting the grains. Recently Tsuneya Ando has reported ballistic transport properties of the CNTs along the tube axis [24]. In the present case it seems that high value of transport $J_{ct}$ arise due to ballistic transport carrying properties of CNTs.

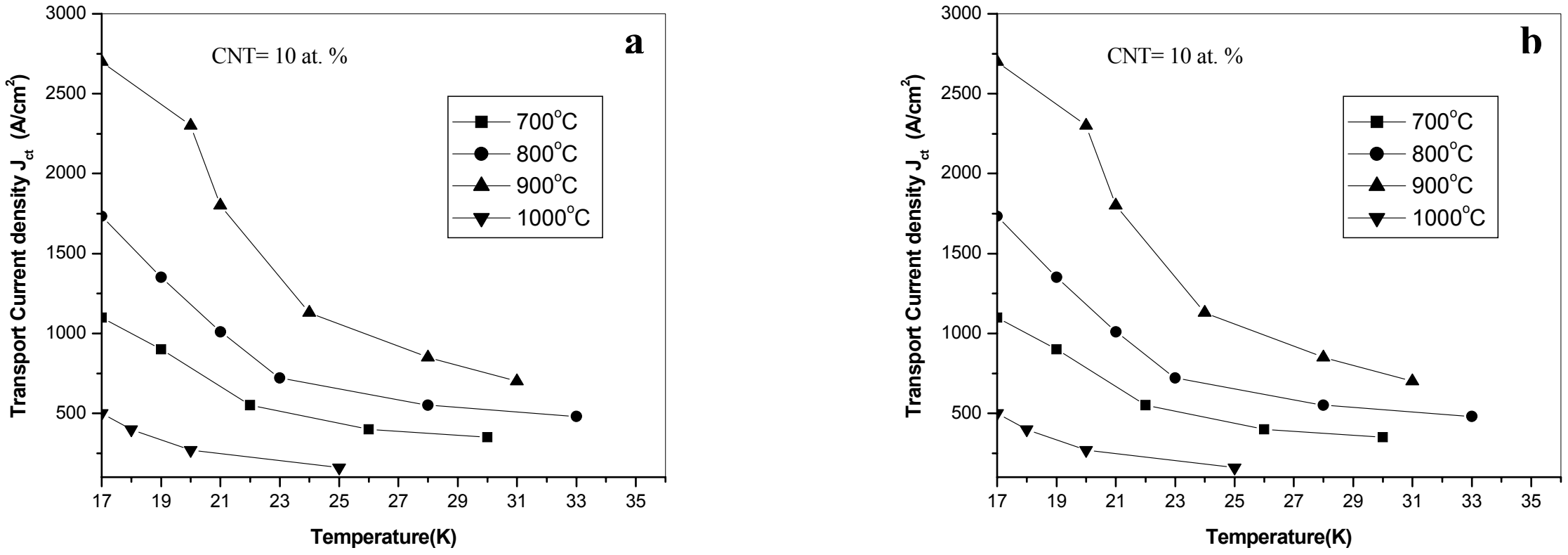

Fig. 3: **(a)** The variation of $J_{ct}$ as a function of temperature for 10 at.% CNTs admixed $MgB_2$ samples sintered at different temperature **(b)** $J_{ct}$ vs temperature behavior of $MgB_2$–x at% CNTs samples ($0 \leq x \leq 25$ at.%) sintered at 900 $^0$C.

The representative surface microstructural feature of $MgB_2$–x at.%CNT composites with CNTs contents of x = 0 at%, 10 at% and 25 at.% are brought out by the SEM micrographs shown in Fig.4a, Fig.4b and Fig.4c respectively. With addition of CNTs in $MgB_2$ the CNTs are not discernable upto x<10 at.%. When concentration of CNTs becomes 10 at.%, the SEM of this sample as compared to the pure sample show that the CNTs is dispersed into $MgB_2$ for the 10 at.% admixed samples (Fig. 4 a&b). Fig 4c shows a typical high density configuration of CNTs in the sample with 25 at.% CNTs. These CNTs in $MgB_2$ matrix may enhance the critical current density due to their special geometry with high aspect ratio (the length and diameter of CNT are ~ 1 to 5 μm and ~5 to 20nm respectively) and possible ballistic transport properties.

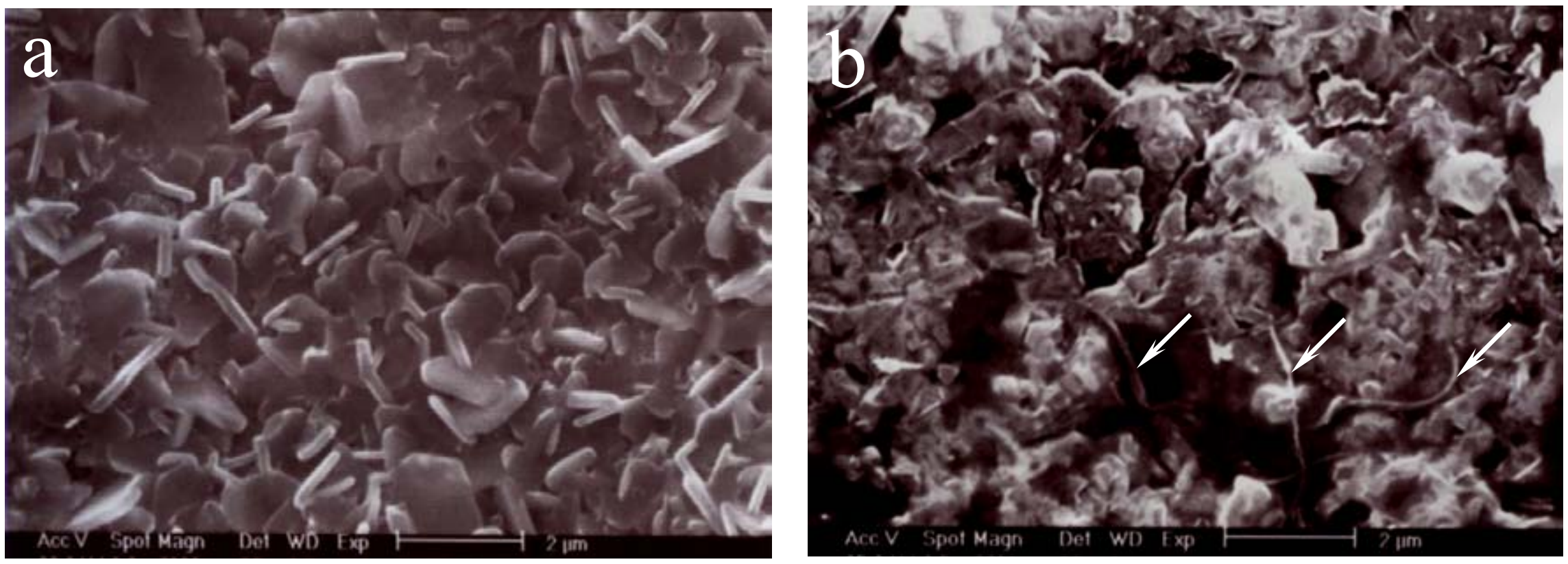

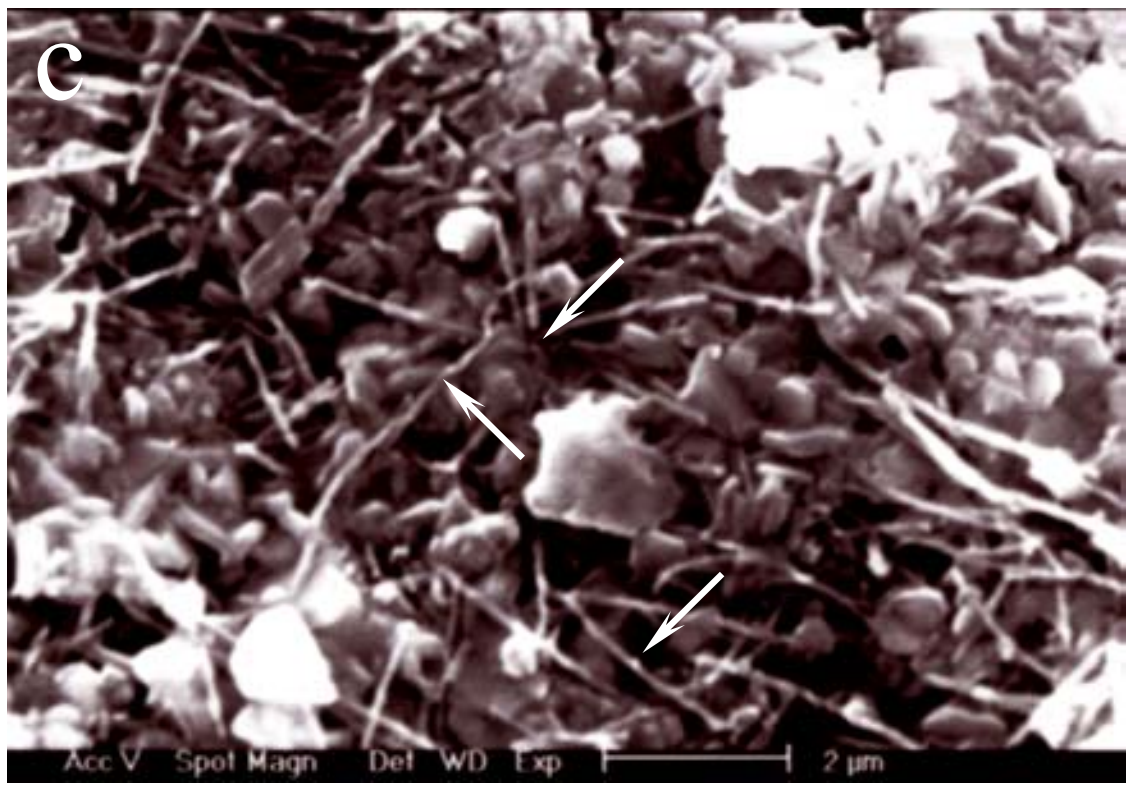

Fig. 4 : A representative SEM micrograph of $MgB_2$–x at% CNTs samples with **(a)** x=0, **(b)** x=10 and **(c)** x=25 at.%. Presence of CNTs in $MgB_2$ matrix (some of which are marked by $\rightarrow$) may be noticed.

Fig. 5a shows the typical TEM image for 10 at.% CNTs admixed $MgB_2$ sample. The TEM micrograph reveals the presence of $MgB_2$–CNTs composite. The CNTs connecting the $MgB_2$ grain boundaries are marked by arrow. Selected area diffraction (SAD) pattern corresponding to Fig. 5a is shown in Fig.5b. SAD pattern shows the hexagonal arrangement of diffraction spots of $MgB_2$ lattice structure and comparatively smeared spots (labeled by arrows) corresponding the CNTs present in the sample.

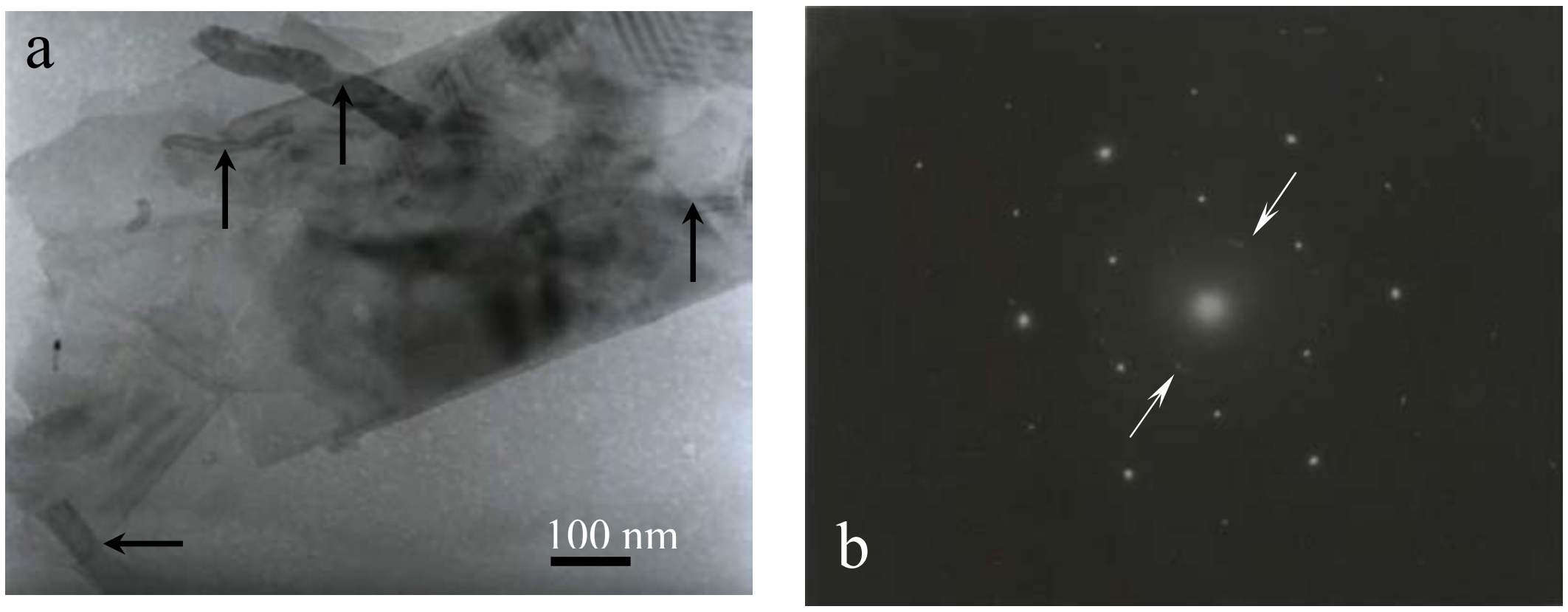

Fig. 5: **(a)** TEM micrograph corresponding to 10 at.% CNTs admixed sample, showing the presence of CNTs connecting the grains of $MgB_2$ (some of which are marked by $\rightarrow$). **(b)** Selected area diffraction pattern from the region shown in (a). Presence of hexagonal arrangement of $MgB_2$ lattice and smeared diffraction spots corresponding to CNTs (marked by $\rightarrow$) may be noticed in SAD pattern.

We have carried out the magnetization measurements as a function of magnetic fields at temperatures 5K, 10K, 20K and 30K, for samples with CNTs contents of x = 0 at.%, 10 at% and 25 at.%. Intragrain critical current density has been calculated from these magnetization measurements using the Bean's formula based on critical state model [22]. The intragrain critical current density ($J_c$) as a function of magnetic field for $MgB_2$, 10 at.% & 25 at.% CNTs admixed $MgB_2$ samples at temperatures of 5K, 10K, 20K and 30K are shown in Fig.6. It is clear from $J_c$ vs H curves, the intragrain $J_c$ of 10 at.% CNTs admixed $MgB_2$ sample attains the highest value amongst all the samples for all temperatures and the whole field region upto 6T. The sample corresponding to 10 at.% CNTs the $J_c$ value at 5K is ~5.2 x$10^6$ A/cm$^2$ in self field, ~1.6 x $10^6$ A/cm$^2$ at 1T, ~2.9 x $10^5$ A/cm$^2$ at 2.6T and ~3.9 x $10^4$ A/cm$^2$ at 4T. While for $MgB_2$ sample (without CNTs) $J_c$ values at 5K are ~1.3$\times10^6$ A/cm$^2$ in self field, ~4.5 $\times10^5$ A/cm$^2$ at 1T, ~7.4 $\times10^4$ A/cm$^2$ at 2.6T and ~1.1 $\times10^4$ A/cm$^2$ at 4T. The $J_c$ for 25 at.% CNTs admixed $MgB_2$ achieves the value of ~2.9 $\times10^6$ A/cm$^2$ and ~5.7 $\times$ $10^5$ A/cm$^2$, ~1.0 $\times10^5$ A/cm$^2$; ~1.9 $\times10^4$ A/cm$^2$ at 5K and in self field, 1T, 2.6T and 4T respectively. These value of $J_c$ are also lower than from 10 at.% CNTs admixed $MgB_2$. Similar variations of $J_c$ with field were also observed for the temperatures 10K, 20K and 30K.

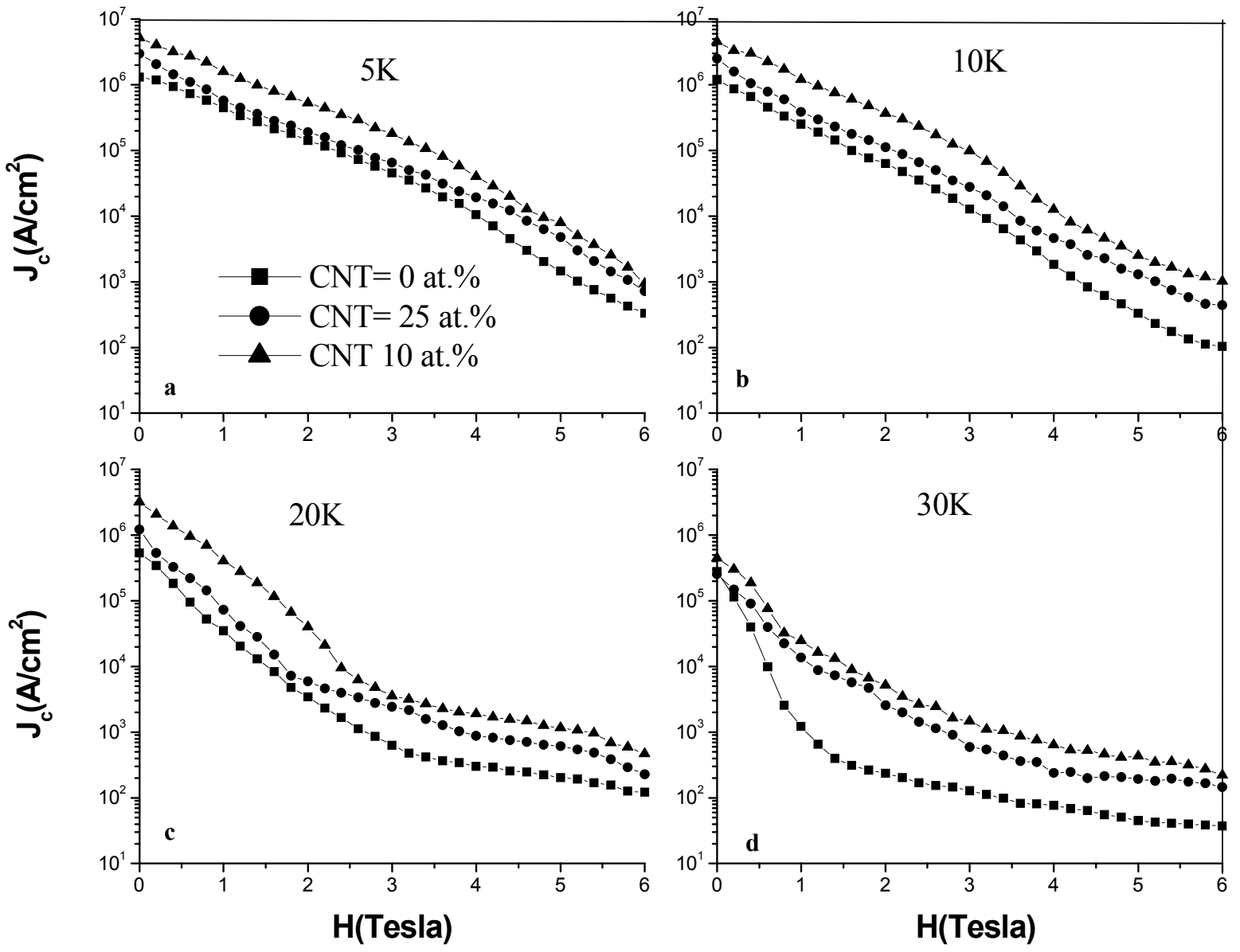

Fig. 6 : Intragrain critical current density (estimated on fine powder version of the samples) as a function of applied magnetic field for $MgB_2$, 10% & 25% CNTs admixed $MgB_2$ samples at temperature 5K, 10K, 20K and 30K. Highest critical current density for the 10 at.% CNTs admixed $MgB_2$ superconductor may be noticed.

In order to find out possible role of CNTs on $J_c$, microstructural explorations of $MgB_2$–x at.% CNTs have been carried out. Figs. 4(b&c) exhibit the dispersion of CNTs in $MgB_2$. As can be seen in the sample $MgB_2$–10 at.% CNTs, there is comparatively uniform distribution of CNTs within grains in the $MgB_2$ matrix. Thus the CNTs can be taken to provide flux pinning center leading to enhacement of $J_c$ from ~$4.5 \times 10^5$ A/cm$^2$ (pure $MgB_2$) to ~$1.6 \times 10^6$ A/cm$^2$ at 1T & 5K (for $MgB_2$–10 at.% CNTs). For the $MgB_2$–25 at% CNT sample, there is a dominance of CNTs (which are also present in the regions outside the grain). This may mask the flux pinning effect. Therefore, enhancement in $J_c$ is not expected to be as high as that for the sample with 10 at.% CNT. The above observations are in keeping with the obtained results as described in the above. It may, therefore, be taken that this is a correlation between the microstructural characteristics and the observed $J_c$ values.

## Conclusion

In conclusion, we have successfully synthesized $MgB_2$–x at.% CNTs (composite material) superconducting samples by solid-state reaction/ sintering at temperatures between ~700-1000 $^0$C at ambient pressure. $T_c$ of the $MgB_2$–x at.% CNTs samples decreases with increasing CNTs content in the samples. The $T_c$ seems to decrease due to substitution of carbon which is present as impurity in the CNTs samples and which substitutes for B. The critical current density of 10 at.% CNTs admixed $MgB_2$ sample attains the highest value amongst all the samples, which corresponds to the comparatively uniform distribution of CNTs in $MgB_2$ matrix. Thus enhancement of critical current density in $MgB_2$–CNTs samples may be attributed to the flux pinning capability of CNTs in $MgB_2$ matrix and better grain connectivity.

## Acknowledgement

The authors are grateful to Prof. A.R. Verma, Prof. C.N.R. Rao, Prof. S.K. Joshi and Prof. T.V. Ramakrishnan for fruitful discussion and suggestions. The authors are also grateful Dr. Anchal Srivastava, Banaras Hindu University Varanasi for his kind support in synthesizing CNTs and fruitful discussions. Financial supports from UGC, DST-UNANST and CSIR are gratefully acknowledged.